# Closed Form Approximations for the Trajectories of the Three Body Problem

AbuBakr Mehmood, Syed Umer Abbas Shah and Ghulam Shabbir

Faculty of Engineering Sciences,

GIK Institute of Engineering Sciences and Technology

Topi, Swabi, NWFP, Pakistan

Email: shabbir@giki.edu.pk

**Abstract:**

The problem of describing the free motions of three gravitating bodies under each other's gravitational influence is one of the oldest of unsolved problems of classical mechanics. Henry Poincare proved in his dictum that due to the nature of the instabilities involved in the problem, it could not be solved. Yet closed form approximations for the problem can be found and it is on these lines that we will explore the problem. It will be shown that closed form approximations for the associated trajectories can be found.

**Introduction:**

In this paper we will develop a procedure to find closed form approximations for the trajectories of three gravitating masses having spherical symmetry, when they perform free motion under each other's gravitational influence. It will be assumed that the masses form an isolated system in free space, and that all the bodies have angular velocities less than 1 $rad/s$ along with bounded position vectors. Other approaches can be found in [1-10]. To begin with, figure 1 shows a typical configuration of the three bodies. It can be shown that the centre of mass this system moves with zero acceleration for all time and hence we do not hesitate in attaching a frame of reference to the centre of mass. Reference frame $oxy$ is therefore an inertial frame of reference.

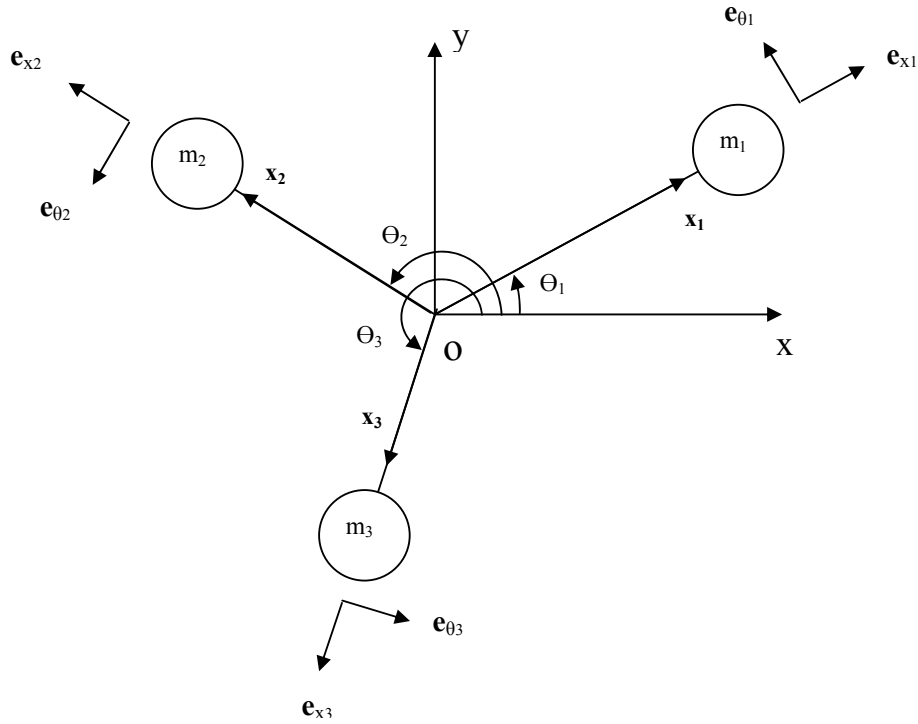

**Figure 1**

Here $\mathbf{x}_i$ is the position vector of a body of mass $m_i$, and $\theta_i$ is the rotation angle of vector $\mathbf{x}_i$ measured form the x axis. In addition $\hat{e}_{x_i}$ is a unit vector in the direction of $\mathbf{x}_i$ and $\hat{e}_{\theta_i}$ is a unit vector perpendicular to $\hat{e}_{x_i}$ where $i = 1, 2, 3$. For the position vectors we will use the notation $\mathbf{x}_i = x_i(t)\hat{e}_{x_i}$ for $i = 1, 2, 3$.

It should be noted that the assumptions of angular velocities being smaller than 1 rad / s and position vectors being bounded are very critical to our derivations. Provided that the position vectors are less than infinity (that is bounded), we can always scale them in a manner such that they become much less than infinity. Stated explicitly, we aim to find approximations for $x_i(\theta_i)$ $\forall$ $\left|\dot{\theta}_i(t)\right| < 1$ rad/s & $x_i(t) \ll \infty$ where $i = 1, 2, 3$.

We also assume that while performing three body motion, each body remains approximately in two body motion with a body having a mass equivalent to the sum of the masses of the other two bodies, placed at their centre of mass. Specifically for each body, our assumption states that

(a)    $m_1$ approximately remains in two body motion with a body of mass $m_2 + m_3$ placed at the centre of mass of $m_2$ and $m_3$ given by

$$\mathbf{x}_{23} = \frac{m_2 \mathbf{x}_2 + m_3 \mathbf{x}_3}{m_2 + m_3} \quad \forall \, t$$

(b)    $m_2$ approximately remains in two body motion with a body of mass $m_1 + m_3$ placed at the centre of mass of $m_1$ and $m_3$ given by

$$\mathbf{x}_{13} = \frac{m_1 \mathbf{x}_1 + m_3 \mathbf{x}_3}{m_1 + m_3} \quad \forall \, t$$

(c)    $m_3$ approximately remains in two body motion with a body of mass $m_1 + m_2$ placed at the centre of mass of $m_1$ and $m_2$ given by

$$\mathbf{x}_{12} = \frac{m_1 \mathbf{x}_1 + m_2 \mathbf{x}_2}{m_1 + m_2} \quad \forall \, t$$

Note that by the use of this assumption, say in (a), we attempt to replicate the individual effects of $m_2$ and $m_3$ on $m_1$ by placing a single body of mass $m_2 + m_3$ at the centre of mass of $m_2$ and $m_3$. However, use of this assumption would turn our solutions into mere approximations instead of exact solutions. Reason being that in order to claim that $m_1$ remains in two body motion with $m_2 + m_3$ located at a point given by $\mathbf{x}_{23}$, we need to show that

$$\hat{e}_{x_1} \cdot \hat{e}_{x_{23}} = -1 \, \forall \, t \tag{1}$$

where $\hat{e}_{x_{23}}$ is a unit vector in the direction of $\mathbf{x}_{23}$. Since the above cannot be shown, it follows that the solution $x_1(\theta_1)$ given by the use of this assumption would be an approximation instead of an exact solution. Stated another way, we are assuming that a body of mass $m_2 + m_3$ located at the centre of mass of $m_2$ and $m_3$ approximately replicates the effects of the two individual bodies $m_2$ and $m_3$, on $m_1$. In accordance, unit vectors $\hat{e}_{x_1}$ and $\hat{e}_{x_{23}}$ remain approximately collinear for all time.

$$\hat{e}_{x_1} \cdot \hat{e}_{x_{23}} \simeq -1 \, \forall \, t \tag{2}$$

Had it been possible to prove relation (1), the three body problem would have been reducible to the two body problem. Similar arguments follow for cases (b) and (c).

For now we consider case (a) and figure 2 presents the configuration of our two body approximation of the three bodies $m_1$, $m_2$ and $m_3$.

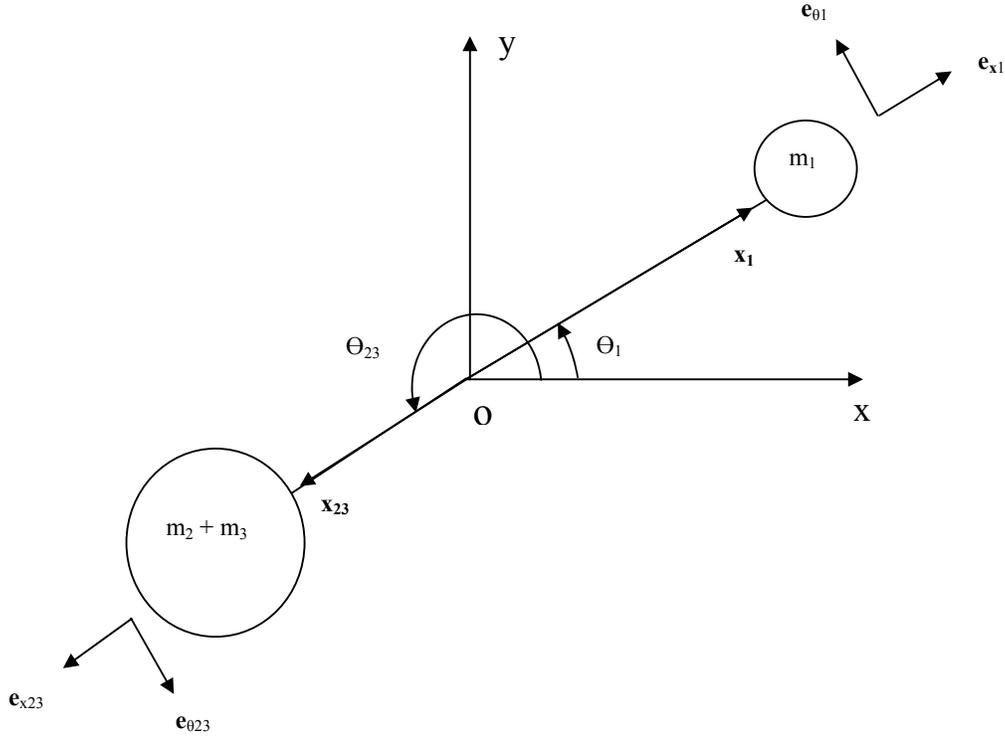

**Figure 2**

In accordance with our assumption, the vectors $\mathbf{x}_1$ and $\mathbf{x}_{23}$ are approximately collinear for all time, and this fact is clearly illustrated in figure 2. This in turn implies that $\theta_1(t) \simeq \theta_{23}(t) + \pi \Rightarrow \dot{\theta}_1(t) \simeq \dot{\theta}_{23}(t)$. Note that in figure 2, $\hat{e}_{x_{23}}$ is a unit vector in the direction of $\mathbf{x}_{23}$ and $\hat{e}_{\theta_{23}}$ is a unit vector perpendicular to $\hat{e}_{x_{23}}$. Modeling this system using Newton's law of Gravitation and Newton's 2nd law, we get

$$\ddot{\mathbf{x}}_1 = -\left( \frac{G(m_2 + m_3)}{|\mathbf{x}_{23} - \mathbf{x}_1|^2} \right) \hat{e}_{x_1} \qquad (3)$$

$$\ddot{\mathbf{x}}_{23} = -\left( \frac{Gm_1}{|\mathbf{x}_{23} - \mathbf{x}_1|^2} \right) \hat{e}_{x_{23}} \qquad (4)$$

Now resolving (3) and (4) in polar coordinates and comparing the coefficients of the unit vectors on both sides of each equation, we can show that the following relations hold true.

$$\ddot{x}_1 - x_1 \dot{\theta}_1^2 = -\left(\frac{G(m_2 + m_3)}{|\mathbf{x}_{23} - \mathbf{x}_1|^2}\right) \tag{5}$$

$$x_1 \ddot{\theta}_1 + 2\dot{x}_1 \dot{\theta}_1 = 0 \tag{6}$$

$$\ddot{x}_{23} - x_{23} \dot{\theta}_{23}^2 = -\left(\frac{Gm_1}{|\mathbf{x}_{23} - \mathbf{x}_1|^2}\right) \tag{7}$$

$$x_{23} \ddot{\theta}_{23} + 2\dot{x}_{23} \dot{\theta}_{23} = 0 \tag{8}$$

It should be noted that a simultaneous solution of equations (5) through (8) would yield $x_1(t)$, $\theta_1(t)$, $x_{23}(t)$ and $\theta_{23}(t)$. We could then make use of the expressions $x_1(t)$ and $\theta_1(t)$ to find $x_1(\theta_1)$, hence describing the trajectory of $m_1$ under the influence of $m_2$ and $m_3$. However, the system of equations (5), (6), (7) and (8) does not seem solvable analytically. We therefore reject the idea of using this approach and adopt an alternate method.

We recall that the problem is being solved for the case $|\dot{\theta}_1(t)| < 1$ rad/s and $x_1(t) \ll \infty$. We could find an approximation to (5) by noting that $\dot{\theta}_1^2 \simeq 0 \ \forall \ t \Rightarrow x_1 \dot{\theta}_1^2 \simeq 0 \ \forall \ t$ (since $|\dot{\theta}_1(t)| < 1$ rad/s and $x_1(t) \ll \infty$). Hence the left hand side of (5) can be approximated as $\ddot{x}_1 - x_1 \dot{\theta}_1^2 \simeq \ddot{x}_1$, yielding the following approximation to (5).

$$\ddot{x}_1 \simeq -\left(\frac{G(m_2 + m_3)}{|\mathbf{x}_{23} - \mathbf{x}_1|^2}\right) \tag{9a}$$

It can be shown that if $|\mathbf{x}_2| \ll \infty$ and $|\mathbf{x}_3| \ll \infty$, then $|\mathbf{x}_{23}| \ll \infty$. Also, if $|\dot{\theta}_1(t)| < 1$ rad/s then $|\dot{\theta}_{23}(t)| < 1$ rad/s (since $\dot{\theta}_1(t) \simeq \dot{\theta}_{23}(t) \ \forall \ t$). Using these arguments, just like we found (9a) to be an approximation to (5), we can find (9b) to be an approximation to (7).

$$\ddot{x}_{23} \simeq -\left(\frac{Gm_1}{|\mathbf{x}_{23} - \mathbf{x}_1|^2}\right) \tag{9b}$$

We now define a vector

$$\mathbf{x} = \mathbf{x}_{23} - \mathbf{x}_1 \tag{10}$$

Since **x** is a vector along the direction of $\mathbf{x}_{23}$, it follows that a unit vector along the direction of **x** is the same as a unit vector along the direction of $\mathbf{x}_{23}$. We define $\hat{e}_x$ to be a unit vector along the direction of **x**. Hence, it follows that $\hat{e}_x = \hat{e}_{x_{23}}$. Note that $\mathbf{x}_1$, $\mathbf{x}_{23}$ and **x** have the representation

$$\mathbf{x} = x(t)\hat{e}_x, \quad \mathbf{x}_1 = x_1(t)\hat{e}_{x_1}, \quad \mathbf{x}_{23} = x_{23}(t)\hat{e}_{x_{23}} \tag{11}$$

and

$$\hat{e}_x = \hat{e}_{x_{23}} = -\hat{e}_{x_1} \tag{12}$$

Making use of (10) and (11), we can write the sum of (9a) and (9b) as

$$\ddot{x}_1(t) + \ddot{x}_{23}(t) = -\left(\frac{G(m_1 + m_2 + m_3)}{x^2(t)}\right) \tag{13}$$

Now use of (11), (12) and (13), along with the definition of **x**, enables us to show that

$$\ddot{x}(t) = -\left(\frac{G(m_1 + m_2 + m_3)}{x^2(t)}\right) \tag{14}$$

We then integrate $(9a)$, $(9b)$ and $(14)$ twice with respect to time and perform a few manipulations on the resulting equations to get

$$x_1(t) = x_{1o} + \dot{x}_{1o} * (t - t_o) + \left(\frac{(m_2 + m_3)}{(m_1 + m_2 + m_3)}\right)[x(t) - \dot{x}_o * (t - t_o) - x_o] \tag{15}$$

where $x_{1o} = x_1(t_o)$, $\dot{x}_{1o} = \dot{x}_1(t_o)$, $x_o = x(t_o)$, $\dot{x}_o = \dot{x}(t_o)$, and $t_o$ is the starting time (normally taken to be zero). In order to find the analytic expression for the trajectory of $m_1$ by use of $(15)$, we must find $t(\theta_1)$. Doing so would help us find $x_1(\theta_1)$ since all variables in $(15)$ would become expressible as functions of $\theta_1$. Therefore we can rewrite $(15)$ as

$$x_1(\theta_1) = x_{1o} - \left(\frac{(m_2 + m_3)}{(m_1 + m_2 + m_3)}\right)x_o + \left[\left(\frac{(m_2 + m_3)}{(m_1 + m_2 + m_3)}\right)\dot{x}_o - \dot{x}_{1o}\right]t_o \tag{16}$$

$$+ \left[\dot{x}_{1o} - \dot{x}_o\left(\frac{(m_2 + m_3)}{(m_1 + m_2 + m_3)}\right)\right]t(\theta_1) + \left(\frac{(m_2 + m_3)}{(m_1 + m_2 + m_3)}\right)x(\theta_1)$$

All that remains to be done to find $x_1(\theta_1)$ explicitly, is to find $t(\theta_1)$ and $x(\theta_1)$ and substitute these expressions into $(16)$. We now begin with the task of finding $t(\theta_1)$ and $x(\theta_1)$. Making use of relations (3), (4), (10), (11) and (12) we can show that the following results hold true.

$$\ddot{x} - x\dot{\theta}^2 = -\left(\frac{G(m_1 + m_2 + m_3)}{x^2}\right) \quad (17)$$

$$x\ddot{\theta} + 2\dot{x}\dot{\theta} = 0 \quad (18)$$

Here '$\theta$' is the rotation angle of vector **x** (which is the same as the rotation angle $\theta_{23}$ of vector $\mathbf{x}_{23}$). Solving equations (17) and (18) simultaneously, it can be shown that

$$x(\theta) = \frac{1}{\left[c_1 \cos\theta + c_2 \sin\theta + \dfrac{G(m_1 + m_2 + m_3)}{x_o^4 \dot{\theta}_o^2}\right]} \quad (19)$$

where $c_1$ and $c_2$ are constants of integration, determinable by incorporation of the initial conditions. Now since $\theta = \theta_{23} \simeq \theta_1 - \pi$, we can write $x(\theta) \simeq x(\theta_1 - \pi)$. Use of this relation enables us to derive the result

$$x(\theta_1) = \frac{1}{\left[k_1 \cos(\theta_1 - \phi_1) + k_2\right]} \quad (20)$$

where $k_1 = \pm\sqrt{c_1^2 + c_2^2}$, $k_2 = \dfrac{G(m_1 + m_2 + m_3)}{x_o^4 \dot{\theta}_o^2}$ and $\phi_1 = \tan^{-1}\left(\dfrac{c_2}{c_1}\right)$. In addition, $t(\theta_1)$ can be shown to satisfy

$$t(\theta_1) = \frac{1}{x_o^2 \dot{\theta}_o} \int_{\theta_{1o}}^{\theta_1} \left[\frac{1}{[k_1 \cos(\theta_1 - \phi_1) + k_2]}\right]^2 d\theta_1 + t_o \quad (21)$$

An evaluation of the integral in the above equation gives

$$t(\theta_1) = \left(\frac{2}{x_o^2 \dot{\theta}_o}\right)\left(\frac{k_1 + k_2}{k_2 - k_1}\right)[(k_1 + k_2)(k_2 - k_1)]^{-\frac{1}{2}}\left[\begin{array}{c}-k_1 \tan^2(0.5\phi - 0.5\theta_1) + k_1 + k_2 \\ +k_2 \tan^2(0.5\phi - 0.5\theta_1)\end{array}\right]^{-1} *$$

$$\left\{\begin{array}{c} k_1((k_1+k_2)(k_2-k_1))^{\frac{1}{2}}\tan(0.5\phi-0.5\theta_1)+k_1\tan^{-1}\left[\dfrac{(k_2-k_1)\tan(0.5\phi-0.5\theta_1)}{((k_1+k_2)(k_2-k_1))^{\frac{1}{2}}}\right] \\ +k_2\tan^2(0.5\phi-0.5\theta_1)-k_1k_2\tan^{-1}\left[\dfrac{(k_2-k_1)\tan(0.5\phi-0.5\theta_1)}{((k_1+k_2)(k_2-k_1))^{\frac{1}{2}}}\right] \\ -k_2^2\tan^2(0.5\phi-0.5\theta_1)\tan^{-1}\left[\dfrac{(k_2-k_1)\tan(0.5\phi-0.5\theta_1)}{((k_1+k_2)(k_2-k_1))^{\frac{1}{2}}}\right] \\ -k_2^2\tan^{-1}\left[\dfrac{(k_2-k_1)\tan(0.5\phi-0.5\theta_1)}{((k_1+k_2)(k_2-k_1))^{\frac{1}{2}}}\right] \end{array}\right\}$$

$$-\left(\dfrac{2}{x_o^2\dot{\theta}_o}\right)\left(\dfrac{k_1+k_2}{k_2-k_1}\right)\left[(k_1+k_2)(k_2-k_1)\right]^{-\frac{1}{2}}\left[\begin{array}{c}-k_1\tan^2(0.5\phi-0.5\theta_{1o})+k_1+k_2 \\ +k_2\tan^2(0.5\phi-0.5\theta_{1o})\end{array}\right]^{-1}*$$

$$\left\{\begin{array}{c} k_1((k_1+k_2)(k_2-k_1))^{\frac{1}{2}}\tan(0.5\phi-0.5\theta_{1o})+k_1\tan^{-1}\left[\dfrac{(k_2-k_1)\tan(0.5\phi-0.5\theta_{1o})}{((k_1+k_2)(k_2-k_1))^{\frac{1}{2}}}\right] \\ +k_2\tan^2(0.5\phi-0.5\theta_{1o})-k_1k_2\tan^{-1}\left[\dfrac{(k_2-k_1)\tan(0.5\phi-0.5\theta_{1o})}{((k_1+k_2)(k_2-k_1))^{\frac{1}{2}}}\right] \\ -k_2^2\tan^2(0.5\phi-0.5\theta_{1o})\tan^{-1}\left[\dfrac{(k_2-k_1)\tan(0.5\phi-0.5\theta_{1o})}{((k_1+k_2)(k_2-k_1))^{\frac{1}{2}}}\right] \\ -k_2^2\tan^{-1}\left[\dfrac{(k_2-k_1)\tan(0.5\phi-0.5\theta_{1o})}{((k_1+k_2)(k_2-k_1))^{\frac{1}{2}}}\right] \end{array}\right\}$$

$$+t_o \qquad (22)$$

Substitution of (20) and (22) into equation (16) would then yield the expression for $x_1(\theta_1)$ as

$$x_1(\theta_1)=x_{1o}-\left(\dfrac{m_2+m_3}{m_1+m_2+m_3}\right)x_o+\left[\left(\dfrac{m_2+m_3}{m_1+m_2+m_3}\right)\dot{x}_o-\dot{x}_{1o}\right]t_o+\left(\dfrac{m_2+m_3}{m_1+m_2+m_3}\right)*$$

$$\left[\dfrac{1}{k_1\cos(\theta_1-\phi_1)+k_2}\right]+\left[\dot{x}_{1o}-\left(\dfrac{m_2+m_3}{m_1+m_2+m_3}\right)\dot{x}_o\right]*$$

$$\left(\frac{2}{x_o^2 \dot{\theta}_o}\right)\left(\frac{k_1+k_2}{k_2-k_1}\right)\left[(k_1+k_2)(k_2-k_1)\right]^{-\frac{1}{2}}\begin{bmatrix}-k_1\tan^2(0.5\phi-0.5\theta_1)+k_1+k_2\\+k_2\tan^2(0.5\phi-0.5\theta_1)\end{bmatrix}*$$

$$\begin{Bmatrix}k_1((k_1+k_2)(k_2-k_1))^{\frac{1}{2}}\tan(0.5\phi_1-0.5\theta_1)+k_1\tan^{-1}\left[\dfrac{(k_2-k_1)\tan(0.5\phi_1-0.5\theta_1)}{((k_1+k_2)(k_2-k_1))^{\frac{1}{2}}}\right]\\[6pt]+k_2\tan^2(0.5\phi_1-0.5\theta_1)-k_1k_2\tan^{-1}\left[\dfrac{(k_2-k_1)\tan(0.5\phi_1-0.5\theta_1)}{((k_1+k_2)(k_2-k_1))^{\frac{1}{2}}}\right]\\[6pt]-k_2^2\tan^2(0.5\phi_1-0.5\theta_1)\tan^{-1}\left[\dfrac{(k_2-k_1)\tan(0.5\phi_1-0.5\theta_1)}{((k_1+k_2)(k_2-k_1))^{\frac{1}{2}}}\right]\\[6pt]-k_2^2\tan^{-1}\left[\dfrac{(k_2-k_1)\tan(0.5\phi_1-0.5\theta_1)}{((k_1+k_2)(k_2-k_1))^{\frac{1}{2}}}\right]\end{Bmatrix}$$

$$-\left(\frac{2}{x_o^2 \dot{\theta}_{1o}}\right)\left(\frac{k_1+k_2}{k_2-k_1}\right)\left[(k_1+k_2)(k_2-k_1)\right]^{-\frac{1}{2}}\begin{bmatrix}-k_1\tan^2(0.5\phi-0.5\theta_{1o})+k_1+k_2\\+k_2\tan^2(0.5\phi-0.5\theta_{1o})\end{bmatrix}*$$

$$\begin{Bmatrix}k_1((k_1+k_2)(k_2-k_1))^{\frac{1}{2}}\tan(0.5\phi_1-0.5\theta_{1o})+k_1\tan^{-1}\left[\dfrac{(k_2-k_1)\tan(0.5\phi_1-0.5\theta_{1o})}{((k_1+k_2)(k_2-k_1))^{\frac{1}{2}}}\right]\\[6pt]+k_2\tan^2(0.5\phi_1-0.5\theta_{1o})-k_1k_2\tan^{-1}\left[\dfrac{(k_2-k_1)\tan(0.5\phi_1-0.5\theta_{1o})}{((k_1+k_2)(k_2-k_1))^{\frac{1}{2}}}\right]\\[6pt]-k_2^2\tan^2(0.5\phi_1-0.5\theta_{1o})\tan^{-1}\left[\dfrac{(k_2-k_1)\tan(0.5\phi_1-0.5\theta_{1o})}{((k_1+k_2)(k_2-k_1))^{\frac{1}{2}}}\right]\\[6pt]-k_2^2\tan^{-1}\dfrac{(k_2-k_1)\tan(0.5\phi_1-0.5\theta_{1o})}{((k_1+k_2)(k_2-k_1))^{\frac{1}{2}}}\end{Bmatrix}$$

$$+t_o$$

(23)

Having found $x_1(\theta_1)$, we go on to find $x_2(\theta_2)$ followed by $x_3(\theta_3)$ following a similar sequence of steps. We do not hesitate in presenting the results directly, reason being that the procedure and all arguments are essentially the same for any pair of gravitating bodies, be it $m_1$ and $m_2+m_3$, $m_2$ and $m_1+m_3$ or $m_3$ and $m_1+m_2$, as stated explicitly in (a), (b) and (c). Figures 3 and 4 present the configurations in accordance with our assumptions (b) and (c).

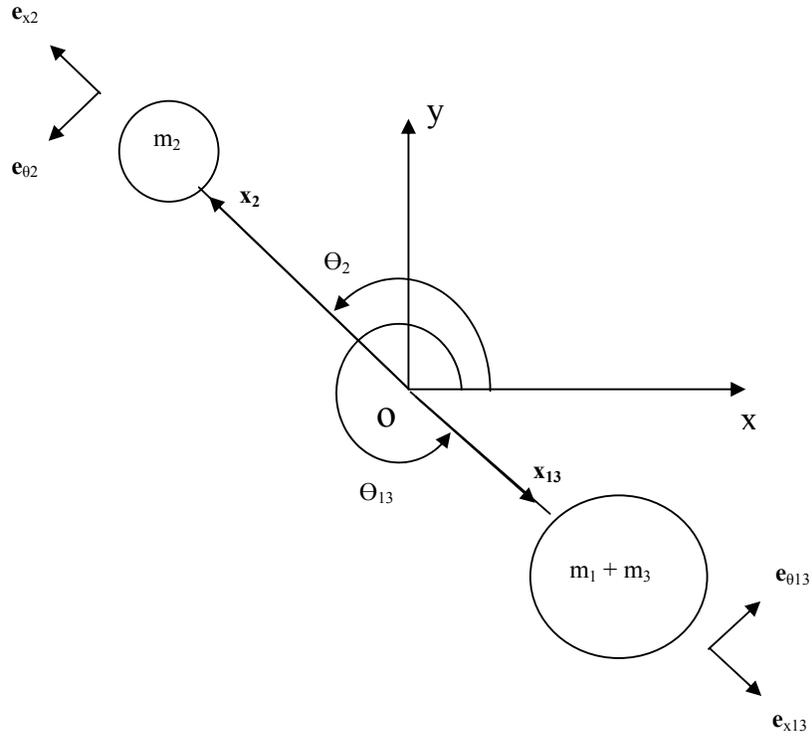

**Figure 3**

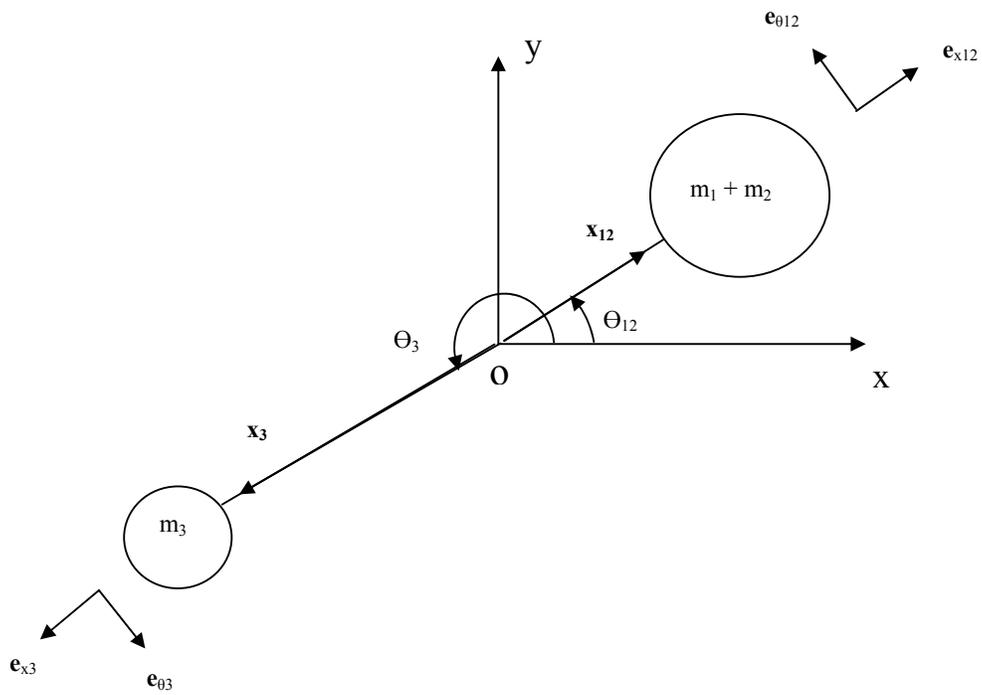

**Figure 4**

It follows that proceeding in exactly the same way as we did for (a) (while finding $x_1(\theta_1)$), we can find $x_2(\theta_2)$ and $x_3(\theta_3)$. It is worth noting that the constant $k_3$ in $x_2(\theta_2)$ and $k_5$ in $x_3(\theta_3)$ are defined in a way analogous to $k_1$ in $x_1(\theta_1)$. Similarly the procedure for defining $k_4$ in case of $x_2(\theta_2)$ and $k_6$ in case of $x_3(\theta_3)$ is exactly analogous to the one for $k_2$ in case of $x_1(\theta_1)$. Same holds true when defining $\phi_2$ and $\phi_3$ in $x_2(\theta_2)$ and $x_3(\theta_3)$ respectively. Having presented the guidelines to be followed, we present equations (24) and (25).

$$x_2(\theta_2) = x_{2o} - \left(\frac{m_1 + m_3}{m_1 + m_2 + m_3}\right) y_o + \left[\left(\frac{m_1 + m_3}{m_1 + m_2 + m_3}\right) \dot{y}_o - \dot{x}_{2o}\right] t_o + \left(\frac{m_1 + m_3}{m_1 + m_2 + m_3}\right) *$$

$$\left[\frac{1}{k_3 \cos(\theta_2 - \phi_2) + k_4}\right] + \left[\dot{x}_{2o} - \left(\frac{m_1 + m_3}{m_1 + m_2 + m_3}\right) \dot{y}_o\right] *$$

$$\left(\frac{2}{\dot{y}_o^2 \theta_{2o}}\right)\left(\frac{k_3 + k_4}{k_4 - k_3}\right)[(k_3 + k_4)(k_4 - k_3)]^{-\frac{1}{2}}\left[\begin{array}{c}-k_3 \tan^2(0.5\phi_2 - 0.5\theta_2) + k_3 + k_4 \\ +k_4 \tan^2(0.5\phi_2 - 0.5\theta_2)\end{array}\right] *$$

$$\left\{\begin{array}{c} k_3((k_3 + k_4)(k_4 - k_3))^{\frac{1}{2}} \tan(0.5\phi_2 - 0.5\theta_2) + k_3 \tan^{-1}\left[\dfrac{(k_4 - k_3)\tan(0.5\phi_2 - 0.5\theta_2)}{((k_3 + k_4)(k_4 - k_3))^{\frac{1}{2}}}\right] \\ \\ +k_4 \tan^2(0.5\phi_2 - 0.5\theta_2) - k_3 k_4 \tan^{-1}\left[\dfrac{(k_4 - k_3)\tan(0.5\phi_2 - 0.5\theta_2)}{((k_3 + k_4)(k_4 - k_3))^{\frac{1}{2}}}\right] \\ \\ -k_4^2 \tan^2(0.5\phi_2 - 0.5\theta_2) \tan^{-1}\left[\dfrac{(k_4 - k_3)\tan(0.5\phi_2 - 0.5\theta_2)}{((k_3 + k_4)(k_4 - k_3))^{\frac{1}{2}}}\right] \\ \\ -k_4^2 \tan^{-1}\left[\dfrac{(k_4 - k_3)\tan(0.5\phi_2 - 0.5\theta_2)}{((k_3 + k_4)(k_4 - k_3))^{\frac{1}{2}}}\right] \end{array}\right\}$$

$$-\left(\frac{2}{\dot{y}_o^2 \theta_{2o}}\right)\left(\frac{k_3 + k_4}{k_4 - k_3}\right)[(k_3 + k_4)(k_4 - k_3)]^{-\frac{1}{2}}\left[\begin{array}{c}-k_3 \tan^2(0.5\phi_2 - 0.5\theta_{2o}) + k_3 + k_4 \\ +k_4 \tan^2(0.5\phi_2 - 0.5\theta_{2o})\end{array}\right] *$$

$$\left\{\begin{array}{c} k_3((k_3 + k_4)(k_4 - k_3))^{\frac{1}{2}} \tan(0.5\phi_2 - 0.5\theta_{2o}) + k_3 \tan^{-1}\left[\dfrac{(k_4 - k_3)\tan(0.5\phi_2 - 0.5\theta_{2o})}{((k_3 + k_4)(k_4 - k_3))^{\frac{1}{2}}}\right] \\ \\ +k_4 \tan^2(0.5\phi_2 - 0.5\theta_{2o}) - k_3 k_4 \tan^{-1}\left[\dfrac{(k_4 - k_3)\tan(0.5\phi_2 - 0.5\theta_{2o})}{((k_3 + k_4)(k_4 - k_3))^{\frac{1}{2}}}\right] \\ \\ -k_4^2 \tan^2(0.5\phi_2 - 0.5\theta_{2o}) \tan^{-1}\left[\dfrac{(k_4 - k_3)\tan(0.5\phi_2 - 0.5\theta_{2o})}{((k_3 + k_4)(k_4 - k_3))^{\frac{1}{2}}}\right] \\ \\ -k_4^2 \tan^{-1}\left[\dfrac{(k_4 - k_3)\tan(0.5\phi_2 - 0.5\theta_{2o})}{((k_3 + k_4)(k_4 - k_3))^{\frac{1}{2}}}\right] \end{array}\right\}$$

$$+ t_o$$

(24)

$$x_3(\theta_3) = x_{3o} - \left(\frac{m_1 + m_2}{m_1 + m_2 + m_3}\right)z_o + \left[\left(\frac{m_1 + m_2}{m_1 + m_2 + m_3}\right)\dot{z}_o - \dot{x}_{3o}\right]t_o + \left(\frac{m_1 + m_2}{m_1 + m_2 + m_3}\right)*$$

$$\frac{1}{k_5 \cos(\theta_3 - \phi_3) + k_6} + \left[\dot{x}_{3o} - \left(\frac{m_1 + m_2}{m_1 + m_2 + m_3}\right)\dot{z}_o\right]*$$

$$\left(\frac{2}{\dot{z}_o^2 \dot{\theta}_{3o}}\right)\left(\frac{k_5 + k_6}{k_6 - k_5}\right)\left[(k_5 + k_6)(k_6 - k_5)\right]^{-\frac{1}{2}}\left[\begin{array}{c}-k_5 \tan^2(0.5\phi_3 - 0.5\theta_3) + k_5 + k_6 \\ +k_6 \tan^2(0.5\phi_3 - 0.5\theta_3)\end{array}\right]*$$

$$\left\{\begin{array}{c}k_5((k_5 + k_6)(k_6 - k_5))^{\frac{1}{2}} \tan(0.5\phi_3 - 0.5\theta_3) + k_5 \tan^{-1}\left[\frac{(k_6 - k_5)\tan(0.5\phi_3 - 0.5\theta_3)}{((k_5 + k_6)(k_6 - k_5))^{\frac{1}{2}}}\right] \\ +k_6 \tan^2(0.5\phi_3 - 0.5\theta_3) - k_5 k_6 \tan^{-1}\left[\frac{(k_6 - k_5)\tan(0.5\phi_3 - 0.5\theta_3)}{((k_5 + k_6)(k_6 - k_5))^{\frac{1}{2}}}\right] \\ -k_6^2 \tan^2(0.5\phi_3 - 0.5\theta_3)\tan^{-1}\left[\frac{(k_6 - k_5)\tan(0.5\phi_3 - 0.5\theta_3)}{((k_5 + k_6)(k_6 - k_5))^{\frac{1}{2}}}\right] \\ -k_6^2 \tan^{-1}\left[\frac{(k_6 - k_5)\tan(0.5\phi_3 - 0.5\theta_3)}{((k_5 + k_6)(k_6 - k_5))^{\frac{1}{2}}}\right]\end{array}\right\}$$

$$-\left(\frac{2}{\dot{z}_o^2 \dot{\theta}_{3o}}\right)\left(\frac{k_5 + k_6}{k_6 - k_5}\right)\left[(k_5 + k_6)(k_6 - k_5)\right]^{-\frac{1}{2}}\left[\begin{array}{c}-k_5 \tan^2(0.5\phi_3 - 0.5\theta_{3o}) + k_5 + k_6 \\ +k_6 \tan^2(0.5\phi_3 - 0.5\theta_{3o})\end{array}\right]*$$

$$\left\{\begin{array}{c}k_5((k_5 + k_6)(k_6 - k_5))^{\frac{1}{2}} \tan(0.5\phi_3 - 0.5\theta_{3o}) + k_5 \tan^{-1}\left[\frac{(k_6 - k_5)\tan(0.5\phi_3 - 0.5\theta_{3o})}{((k_5 + k_6)(k_6 - k_5))^{\frac{1}{2}}}\right] \\ +k_6 \tan^2(0.5\phi_3 - 0.5\theta_{3o}) - k_5 k_6 \tan^{-1}\left[\frac{(k_6 - k_5)\tan(0.5\phi_3 - 0.5\theta_{3o})}{((k_5 + k_6)(k_6 - k_5))^{\frac{1}{2}}}\right] \\ -k_6^2 \tan^2(0.5\phi_3 - 0.5\theta_{3o})\tan^{-1}\left[\frac{(k_6 - k_5)\tan(0.5\phi_3 - 0.5\theta_{3o})}{((k_5 + k_6)(k_6 - k_5))^{\frac{1}{2}}}\right] \\ -k_6^2 \tan^{-1}\left[\frac{(k_6 - k_5)\tan(0.5\phi_3 - 0.5\theta_{3o})}{((k_5 + k_6)(k_6 - k_5))^{\frac{1}{2}}}\right]\end{array}\right\}$$

$$+t_o$$

(25)

It should be noted that the following two definitions find use in our results (24) and (25). For derivation of (24), we defined $\mathbf{y} = \mathbf{x}_{13} - \mathbf{x}_2$, and similarly, in case of (25), we utilized $\mathbf{z} = \mathbf{x}_{12} - \mathbf{x}_3$. Having derived all the required results, we now go on to summarize our accomplishments.

**Summary:**

We aimed at deriving closed form approximations for the associated trajectories when three masses execute free motion under each other's gravitational influence. The masses were defined to form an isolated system in free space. We assumed that each mass possesed a spherically symmetric mass distribution and that all the bodies had angular velocities less than 1 $rad/s$ along with bounded position vectors. Also, we utilized a single body to replicate the effects of individual bodies in each body. We were able to derive expressions defining the trajectories of these masses in equations (23), (24) and (25). It is worth stating that our assumptions were both practical and feasible. Celestial bodies do posses mass distributions that are approximately spherically symmetric, and angular velocities encountered in celestial motion are doubtlessly less than 1 $rad/s$ (excluding exceptions). Also, we assumed that position vectors remain bounded $(|\mathbf{x}_k| < \infty \; \forall \; k = 1, \; 2 \text{ and } 3)$. Using reasonable units of length, this condition can be satisfied at the scale of solar systems as well as galaxies. Furthermore, we can always scale the position vectors along their length so that the relation $|\mathbf{x}_k| << \infty$ is satisfied for $k = 1, \; 2$ and $3$. Thus, our method is flexible enough to provide us with approximations for the trajectories at the scale of both solar systems and galaxies. The sources of error lie in the fact that we use a single mass to replicate the effect of two individual masses. Nevertheless our aim was to find approximations, not exact solutions.